\documentclass[apj,twocolumn]{openjournal}

\usepackage{xcolor}
\usepackage{textgreek}
\usepackage[utf8]{inputenc}
\usepackage[english]{babel}

\usepackage{hyperref}
\hypersetup{
    unicode, 
    colorlinks=true,
    linkcolor=linkcolor,
    citecolor=linkcolor,
    filecolor=linkcolor,
    urlcolor=linkcolor,
}
\usepackage{color,colortbl}
\definecolor{linkcolor}{rgb}{0.0,0.3,0.5}
\usepackage{tensind}
\tensordelimiter{?}
\DeclareGraphicsExtensions{.bmp,.png,.jpg,.pdf}
\usepackage{verbatim}
\usepackage{placeins}
\usepackage[normalem]{ulem}
\usepackage{orcidlink}
\usepackage{soul}
\usepackage[nolist,nohyperlinks]{acronym}
\newcommand{\snr}{N\,132D}
\newcommand{\cxo}{{\it Chandra}}

\newcommand{\ergcm}[1]{erg\,cm$^{-2}$\,s$^{-1}$}

\newcommand{\OIII}{[O\,{\sc iii}]}

\newcommand{\D}{$^\circ$}
\newcommand{\hour}{$^{\mathrm{h}}$}
\newcommand{\minute}{$^{\mathrm{m}}$}
\newcommand{\second}{$^{\mathrm{s}}$}

\graphicspath{ {./figs/} }

\begin{document}
\title{Radio Continuum Study of One of the Brightest Large Magellanic Cloud Supernova Remnants \snr}

\author{Rami Z. E. Alsaberi\orcidlink{0000-0001-5609-7372}}
\email{ramy\_z@yahoo.com}
\affiliation{Western Sydney University, Locked Bag 1797, Penrith NSW 2751, Australia}
\affiliation{Faculty of Engineering, Gifu University, 1-1 Yanagido, Gifu 501-1193, Japan}

\author{Zachary J. Smeaton\orcidlink{0009-0009-7061-0553}}
\affiliation{Western Sydney University, Locked Bag 1797, Penrith NSW 2751, Australia}

\author{Miroslav D. Filipovi\'c\orcidlink{0000-0002-4990-9288}}
\affiliation{Western Sydney University, Locked Bag 1797, Penrith NSW 2751, Australia}

\author{Roland Kothes\orcidlink{0000-0001-5953-0100}}
\affiliation{Dominion Radio Astrophysical Observatory, Herzberg Astronomy \& Astrophysics, National Research
Council Canada, Penticton, BC, Canada}

\author{Hidetoshi Sano\orcidlink{0000-0003-2062-5692}}
\affiliation{Faculty of Engineering, Gifu University, 1-1 Yanagido, Gifu 501-1193, Japan}
\affiliation{Center for Space Research and Utilization Promotion (c-SRUP), Gifu University, 1-1 Yanagido, Gifu 501-1193, Japan}

\author{Ahmed H. Abdullah\orcidlink{0000-0003-1165-7924}}
\affiliation{Department of Astronomy and Space, College of Science, University of Baghdad, Baghdad, Iraq}

\author{Dejan Uro\v sevi\'c\orcidlink{0000-0003-0665-0939}}
\affiliation{Department of Astronomy, Faculty of Mathematics, University of Belgrade, Studentski trg 16, 11000 Belgrade, Serbia} 
\affiliation{Isaac Newton Institute of Chile, Yugoslavia Branch}

\author{Sanja Lazarevi\'c\orcidlink{0000-0001-6109-8548}}
\affiliation{Western Sydney University, Locked Bag 1797, Penrith NSW 2751, Australia}
\affiliation{Australia Telescope National Facility, CSIRO, Space and Astronomy, PO Box 76, Epping, NSW 1710, Australia}
\affiliation{Astronomical Observatory, Volgina 7, 11060 Belgrade, Serbia }


\begin{abstract}
    We present a new radio continuum study of the Large Magellanic Cloud supernova remnant \snr\ using observations from the \ac{ATCA} and MeerKAT. The radio morphology shows a clear spatial correlation with the X-ray emission in the central and left regions of the remnant, as well as in the right portion of the shell, indicating a close connection between the thermal and non-thermal plasma components. Polarisation measurements reveal predominantly radial magnetic fields, with mean fractional polarisations of 1.2\,per\,cent, 3.0\,per\,cent, and 4.0\,per\,cent at 2100, 5500, and 9000\,MHz, respectively. There is also a complex structure in the Faraday Depth spectrum, showing a double-peaked profile that indicates regions with differing magnetisation properties or varying magnetic field geometry. The rotation measure derived from \ac{ATCA} data has an average value of $\sim2 $\,rad\,m$^{-2}$, with a maximum of $\sim338$\,rad\,m$^{-2}$. We estimate a constant partition magnetic field of $\sim$90\,$\mu$G and a minimum energy of $3\times10^{50}$\,erg, which is in a constant partition between the energy density of electrons and the energy density of the magnetic field. The integrated radio spectrum follows a power law with a spectral index of $-0.74 \pm 0.01$. These results provide new constraints on the magnetic field structure and energetic properties of \snr\ and offer insight into particle acceleration and shock evolution in young supernova remnants.
\end{abstract}


\maketitle

\section{Introduction}

\begin{acronym}
\acro{2MASS}{The Two Micron All-Sky Survey}
\acro{30Dor}[30 Dor]{30~Doradus}
\acro{ALMA}{Atacama Large Millimeter Array}
\acro{AGN}{active galactic nucleus}
\acrodefplural{AGN}{Active Galactic Nuclei}
\acro{ATCA}{Australia Telescope Compact Array}
\acro{ATNF}{Australia Telescope National Facility}
\acro{ATOA}{Australia Telescope Online Archive}
\acro{AT20G}{Australia Telescope 20 GHz Survey}
\acro{ASKAP}{Australian Square Kilometre Array Pathfinder}
\acro{BL}[BL-Lac]{BL Lacertae Objects}
\acro{CASS}{CSIRO Astronomy and Space Science}
\acro{CABB}{Compact Array Broadband Back-end}
\acro{CHII}[{\sc CHii}]{Compact \textsc{Hii}}
\acro{CSIRO}{Australian Commonwealth Scientific and Industrial Research Organisation}
\acro{CSS}{Compact Steep Spectrum}
\acro{CME}{Coronal Mass Ejection}
\acro{CR}{Cosmic Ray}
\acro{Dec}{Declination}
\acro{DD}{double-degenerate}
\acro{DS9}[\textsc{DS9}]{\textsc{SAOImage DS9}}
\acro{DSS}{Digital Sky Survey}
\acro{EMU}{Evolutionary Map of the Universe}
\acro{ev}[eV]{electronvolt\acroextra{: 1 eV $\approx 1.6 \times 10^{-19}$ J}}
\acro{FITS}[\textsc{Fits}]{Flexible Image Transport System}
\acro{FRB}{Fast Radio Bursts}
\acro{FRII}{Fanaroff-Riley II}
\acrodefplural{FRII}{Fanaroff-Riley II's}
\acro{FRI}{Fanaroff-Riley I}
\acrodefplural{FRI}{Fanaroff-Riley I's}
\acro{FSRQ}{Flat Spectrum Radio Quasars}
\acro{FWHM}{Full Width at Half-Maximum}
\acro{GLEAM}{GaLactic Extragalactic All-sky MWA}
\acro{GMRT}{Giant Meterwave Radio Telescope}
\acro{GRG}{Giant Radio Galaxy}
\acro{GPS}{Gigahertz Peak Spectrum}
\acro{HEMT}{High Electron Mobility Transistor}
\acro{HFP}[HFP]{High Frequency Peaker}
\acro{HzRGs}{High Redshift Radio Galaxies}
\acro{HVS}{hypervelocity stars}
\acro{pHFP}[pHFP]{Potential High Frequency Peaker}
\acro{HST}{\textit{Hubble Space Telescope}}
\acro{IAU}{International Astronomical Union}
\acro{IFRSs}{Infrared Faint Radio Sources}
\acro{ISM}{Interstellar Medium}
\acro{IR}{Infrared}
\acro{JY}[Jy]{Jansky\acroextra{, 1 Jy = $10^{-26} \times \mathrm{W~ m}^{-2}~\mathrm{Hz}^{-1}$}} 
\acro{LLS}{Largest Linear Size}
\acro{LFAA}{Low-Frequency Aperture Array}
\acro{LMC}{Large Magellanic Cloud}
\acro{LG}{Local Group}
\acro{LSO}[LSOs]{Large Scale Objects}
\acro{MACHO}{Massive Astrophysical Compact Halo Objects}
\acro{MC}[MCs]{Magellanic Clouds}
\acro{mc}[MC]{Magellanic Cloud}    
\acro{MCELS}{Magellanic Cloud Emission Line Survey}
\acro{MW}{Milky Way}
\acro{MIRIAD}[\textsc{Miriad}]{Multichannel Image Reconstruction, Image Analysis and Display}
\acro{MIT}{Massachusetts Institute of Technology}
\acro{MOST}{Molonglo Observatory Synthesis Telescope}
\acro{MQS}{Magellanic Quasars Survey}
\acro{MRC}{Molonglo Reference Catalogue of Radio Sources}
\acro{MWA}{Murchison Widefield Array}
\acro{NAT}{Narrow-Angle Tail}
\acro{NRAO}{National Radio Astronomy Observatory}
\acro{NVSS}{NRAO VLA Sky Survey}
\acro{ORCs}{Odd Radio Circles}
\acro{ORC}{Odd Radio Circle}
\acro{OPAL}{Online Proposal Applications \& Links}
\acro{OVV}{Optically Violent Variable Quasars}            
\acro{PAF}{Phased Array Feed}
\acro{pc}{parsec\acroextra{: 1 pc $\simeq 3.09 \times 10^{16}$ m}}
\acro{PMN}{Parkes-MIT-NRAO}
\acro{PNe}{Planetary Nebulae}
\acro{PN}{Planetary Nebula}
\acro{PWN}{Pulsar Wind Nebula}
\acro{QSO}{Quasi-Stellar Object}
\acro{RA}{Right Ascension}
\acro{RFI}{Radio-Frequency Interference}
\acro{RMS}[rms]{Root Mean Squared}
\acro{SDSS}{Sloan Digital Sky Survey}
\acro{SED}{spectral energy distribution}
\acro{SI}[$\alpha$]{Spectral Index\acroextra{, $S \propto \nu^\alpha$}}
\acro{SKA}{Square Kilometre Array}
\acro{SMASH}{Survey of the MAgellanic Stellar History}
\acro{SMB}{Super Massive Blackholes}
\acro{SMC}{Small Magellanic Cloud}
\acro{SN}{Supernova}
\acro{SNe}{Supernovae}
\acro{SNR}{Supernova Remnant}
\acrodefplural{SNRs}{Supernova Remnants}
\acro{SD}{single-degenerate}
\acro{SUMSS}{Sydney University Molonglo Sky Survey}
\acro{SMBH}{Super Massive Black Hole}
\acro{TOPCAT}[\textsc{Topcat}]{Tool for OPerations on Catalogues And Tables}
\acro{USS}{Ultra Steep Spectrum}
\acro{WBAC}{Wide-Band Analogue Correlator}
\acro{WIFES}[WiFeS]{Wide-Field Spectrograph}
\acro{WISE}{Wide-Field Infrared Survey Explorer}
\acro{VLA}{Very Large Array} 
\acro{VLBI}{Very Long Baseline Interferometry} 
\acro{VLSR}[\textbf{$v_{lsr}$}]{Velocity in the Line of Sight}
\acro{WD}{white dwarf}
\end{acronym}

\acp{SNR} play an important role in the structure of galaxies, enriching the \ac{ISM} with heavy elements as well as having a significant impact on the structure and physical properties of the \ac{ISM} \citep{book1,book2}. The study of \acp{SNR} in our own Galaxy is not ideal because of the high level of absorption, in addition to the difficulties in achieving accurate distance measurements to objects. Instead, we look into nearby, small, dwarf galaxies such as the \ac{LMC}, which is located at a distance of 50\,kpc \citep{mt,2019Natur.567..200P}. This proximity allows observers to assume that all objects within the galaxy are located at the same distance, making physical measurements, such as the physical extent and surface brightness, more reliable. Additionally, the \ac{LMC} is in a nearly face-on orientation (inclination angle of 35$^{\circ}$) and is near enough to allow deep, high-resolution (spatial and spectral), multi-frequency observations \citep{2016A&A...585A.162M,2017ApJS..230....2B,2023MNRAS.518.2574B}. The \ac{LMC} also contains active star-forming regions and is located outside of the Galactic Plane, where absorption by gas and dust is reasonably low.

\snr\ was first identified as an \ac{SNR} by \cite{1966MNRAS.131..371W}. It is a bright X-ray/radio source \citep{2018ApJ...854...71B,1995AJ....109..200D} which is also known as very high energy ($>$100\,Gev) gamma-ray source \citep{2021A&A...655A...7H}. The age of this \ac{SNR} is estimated at $\sim2500$\,years \citep{2011Ap&SS.331..521V}. \cite{2020ApJ...894...73L} estimated the average expansion velocity for this remnant as 1745\,km\,s$^{-1}$ with an age of $2450\pm195$\,years. Thus, it has been suggested that this \ac{SNR} is transitioning into a middle-aged \ac{SNR} \citep{2018ApJ...854...71B}. \cite{2011ApJS..197...16W} reported the discovery of a CO cloud towards the southern edge of \snr\ using Mopra data, while \cite{2020ApJ...902...53S} inferred a detection of diffuse CO emission inside the X-ray shell using \ac{ALMA} data. 

Radio emission from \acp{SNR} typically emerges as synchrotron emission from the outer shock front. This makes it an effective tracer of the full shell structure, and high-resolution radio-continuum observations are important tools in analysing the full extent of \acp{SNR}, as well as mapping asymmetries and environment interactions. In the literature, we found only one radio continuum study that investigated \snr. This study was presented by \cite{1995AJ....109..200D}. They used \ac{ATCA} observations to report a spectral index value of --0.7, an average rotation measure for the whole remnant of 140\,rad\,m$^{-2}$, and an average fractional polarisation of 4\% and 6\% for 4790 and 8600\,MHz, respectively.

Here we present a new \ac{ATCA} (2100, 5500, and 9000\,MHz) and MeerKAT (1295\,MHz) study for \snr. The structure of the paper is as follows. In Section~\ref{obs}, we illustrate our observations and data analysis. Our findings and discussion are presented in Section~\ref{rus}. Finally, the conclusions are outlined in Section~\ref{con}.

\section{Observations and Analysis} 
\label{obs}

\subsection{Radio Observations}

\subsubsection{ATCA Observations}
Our new \ac{ATCA} observations, including observing dates, project numbers, observation frequency, the time spent on the source, and the array configuration, are presented in Table~\ref{tab:summary_obs}. All observations were carried out in `snap-shot' mode, with 1 hour of integration over a 12-hour period as a minimum. All observations used the \ac{CABB} (with 2048\,MHz bandwidth), centred at wavelengths of 3/6\,cm ($\nu$~=~4500--6500 and 8000--10000\,MHz; centred at 5500 and 9000\,MHz, respectively) totalling 162\,minutes of integration; and 13\,cm ($\nu$~=~1100--3100\,MHz; centred at 2100\,MHz) totalling $\sim$698\,minutes of integration. 


We used \textsc{miriad}\footnote{\url{http://www.atnf.csiro.au/computing/software/miriad/}} \citep{1995ASPC...77..433S}, \textsc{karma}\footnote{\url{http://www.atnf.csiro.au/computing/software/karma/}} \citep{1995ASPC...77..144G}, and \textsc{carta} \citep{carta} software packages for reduction and analysis. All observations were calibrated using the phase and flux calibrators listed in Table~\ref{tab:summary_obs}. Imaging was completed using the multi-frequency synthesis \textsc{invert} task with Briggs weighting (robust~=~--0.5) and a beam size of $4\times4$\,arcsec$^{2}$ for 2100\,MHz and $3\times3$\,arcsec$^{2}$ for 5500 and 9000\,MHz. The \textsc {mfclean} and \textsc {restor} algorithms were used to deconvolve the images, with primary beam correction applied using the \textsc{linmos} task. We follow the same process with Stokes {\it Q} and {\it U} parameters to produce polarisation maps, but with a beam size of $5\times5$\,arcsec$^{2}$ (see Section~\ref{pol} for more details).

We excluded the 6A array data from the \ac{ATCA} imaging at 5500 and 9000\,MHz because the lack of short-spacing resulted in significant missing flux.


\begin{table*}[t]
	\centering
	\caption{Summary of our \ac{ATCA} observations for \snr\ used in this study.}
	\label{tab:summary_obs}
	\begin{tabular}{@{}clccllllcc@{}}
		\hline
    Observing   & Project& Array    & No.       & Bandwidth &  Frequency $\nu$  & Phase         & Flux              & Integrated time  \\
    Date        & Code   & Configuration  & Channels  &  (GHz)    &  (MHz)        & Calibrator        & Calibrator        & (minutes) \\
		\hline
      
    02~Jan~2015 & CX310 & 6A         & 2049      &  2.048    &   2100        & PKS\,B0530--727 	            & PKS\,B1934--638  & 60      \\ 
    26~Mar~2018 & C3229 & EW352      & 2049      &  2.048    &   5500,9000   & PKS\,B0530--727 	             & PKS\,B1934--638  & 59.4   \\
    26~Mar~2018 & C3229 & EW352      & 2049      &  2.048    &   2100        & PKS\,B0530--727 	             & PKS\,B1934--638  & 408.6  \\
    26~Feb~2021 & C3403 & 6D         & 2049      &  2.048    &   5500,9000   & PKS\,B0530--727               & PKS\,B1934--638  & 102.6 \\
    26~Feb~2021 & C3403 & 6D         & 2049      &  2.048    &   2100        & PKS\,B0530--727               & PKS\,B1934--638  & 229.8 \\
		\hline

        \hline\hline
	\end{tabular}
    \vspace{4mm}
\end{table*}

\subsubsection{MeerKAT Observations}

We also use MeerKAT observations from the MeerKAT survey of the \ac{LMC}. These observations were conducted between Aug and Nov 2019 using the L-band receiver over the frequency band 856--1712\,MHz. The data consist of a broadband image over the entire bandwidth centred at a reference frequency of 1295.05\,MHz, as well as 12 sub-band images. 2 of these 12 sub-bands were blanked due to interference. These images were created for all Stokes parameters $I$, $Q$, $U$, and $V$. More details of these observations will be provided in the data release paper \citep{2026MNRAS.549ag963C}. 

We use the broadband Stokes~$I$ image for flux density measurements in the spectral index calculation, and the Stokes~$Q$ and $U$ sub-band images for rotation measure synthesis. All images were resolved to a common beam size of $8\times8$\,arcsec$^{2}$.

The MeerKAT $Q$ and $U$ data were used to perform rotation measure synthesis and generate zeroth moment (M$_0$) and first moment (M$_1$) maps for \snr. This process involved calculating the Faraday Depth function for each pixel, similar to the process used by \citet{2023MNRAS.524.1396B,2025PASA...42..104F} and described in \citet{2005A&A...441.1217B}. The Faraday Depth function was initially calculated over a broad range with a coarse step size (--1500 to +1500 rad m$^{-2}$; step size 50 rad m$^{-2}$) to identify regions of significant emission. The analysis was then refined over a smaller range of --500 to +500 rad m$^{-2}$ with a finer step of 10 rad m$^{-2}$, where the majority of the Faraday structure was found to be present.

The M$_0$ moment map represents the total polarised intensity integrated over all Faraday depths, and the M$_1$ map is the intensity-weighted mean Faraday Depth, defined as:

\begin{equation}
\label{M1eq}
M_1 = \frac{\Sigma\phi_i|F(\phi_i)|^2}{\Sigma|F(\phi_i)|^2},
\end{equation}

where $F(\phi_i)$ is the Faraday Depth function at Faraday Depth $\phi_i$. 

\subsection{Optical Observations}
The \OIII\ \ac{HST} image of the \snr\ was downloaded from the Mikulski Archive for Space Telescopes\footnote{\url{https://mast.stsci.edu/portal/Mashup/Clients/Mast/Portal.html}}. The data were acquired with the HST ACS/WFC using the F502N narrowband filter as part of program 12001 (PI: J. Green) in January~2004. The final drizzled image combines four exposures with a total integration time of approximately 1300--1600~seconds.

\subsection{X-ray Observations}
We utilised archival X-ray observations from \cxo\ X-ray Observatory, corresponding to observation IDs 5532, 7259, and 7266 (PI: K. J. Borkowski; proposal No. 06500305). All details regarding the data reduction for these observations are described in \cite{2020ApJ...902...53S}.

\section{Results and discussion} \label{rus}

\subsection{Morphology}
\begin{figure*}
\begin{center}
\includegraphics[width=\textwidth,trim=0 0 0 0,clip]{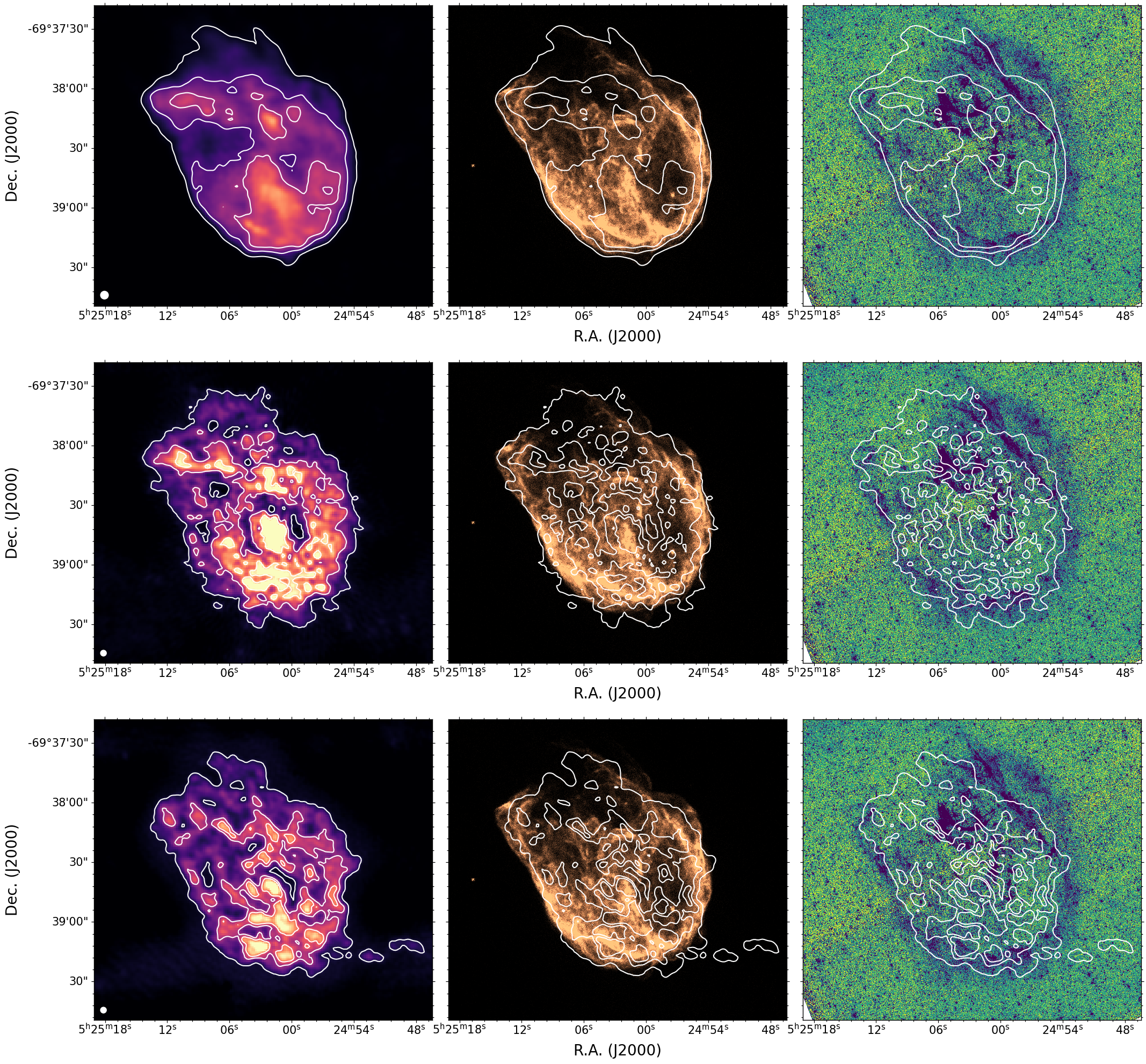}
\caption{Images of \snr\ with 2100\,MHz contours overlaid (top), 5500\,MHz contours overlaid (middle), and 9000\,MHz contours overlaid (bottom). \textbf{Left:} \ac{ATCA} intensity image at 2100\,MHz (top), 5500\,MHz (middle), and 9000\,MHz (bottom). The white circles in the lower left corner represent the synthesised beams: $4\times4$\,arcsec$^{2}$ at 2100\,MHz, and $3\times3$\,arcsec$^{2}$ at both 5500 and 9000\,MHz. \textbf{Middle:} Broadband \cxo\ X-ray images. \textbf{Right:} HST \OIII\ filter images, where darker regions correspond to higher emission, while brighter regions indicate lower emission.} \label{Fig1}
\end{center}
\end{figure*}

\snr\ shows a well-defined shell-like morphology in the radio continuum at 2100, 5500, and 9000\,MHz (Figure~\ref{Fig1}, left panels), with bright emission concentrated along the outer boundary and significant internal structure. The lower-frequency image highlights the diffuse large-scale emission, while the higher-frequency maps reveal finer filamentary features and localised intensity enhancements within the shell. White contours trace the radio emission and emphasise the remnant's overall morphology.

In Figure~\ref{Fig1} we present a comparison between our new \ac{ATCA} images of \snr\ and the corresponding \cxo\ and \OIII\ \ac{HST} images. To investigate the spatial correlation between the radio, optical, and X-ray emission, we extracted azimuthally resolved radial intensity profiles from the geometric centre of the \ac{SNR} (RA(J2000)~=~05\hour25\minute03.1\second, Dec~(J2000)~=~--69\D38$'$22.3$''$) out to 90\,arcsec (1.5\,arcmin). The images were regridded to a common pixel grid and convolved to the same angular resolution. Profiles were sampled at 10\D intervals in position angle, with the intensity averaged over a 5$^{\circ}$ angular width. The resulting profiles were grouped into four 90\D regions: north (315\D--45\D), east (45\D--135\D), south (135\D--225\D), and west (225\D--315\D), with nine position-angle profiles contributing to each region. The mean profiles were independently normalised to their maximum intensities, and the Pearson correlation coefficient was used to quantify their radial correlation. The radius of maximum intensity was also determined for each wavelength and region.

\begin{figure*}
\begin{center}
\includegraphics[width=\textwidth,trim=0 0 0 0,clip]{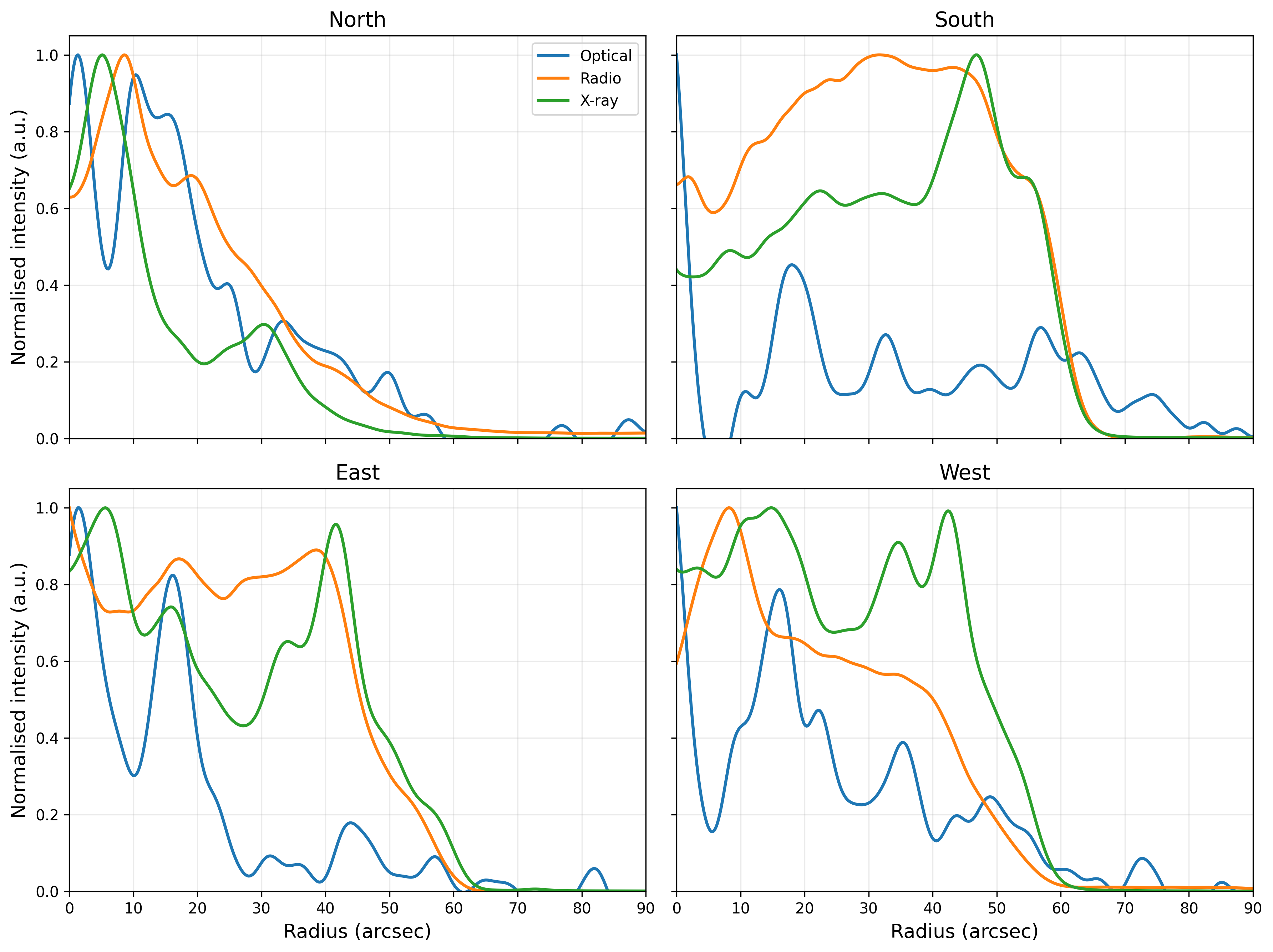}
\caption{Normalised radial intensity profiles of the SNR derived from the optical (\OIII), radio (2100\,MHz), and X-ray (broadband) images for the north, east, south, and west regions, extending to 90\,arcsec from the adopted geometric centre (RA~(J2000)~=~05\hour25\minute03.1\second, Dec~(J2000)~=~--69\D38$'$22.3$''$).}\label{Fig_3}
\end{center}
\end{figure*}

The radio and X-ray profiles show a consistently strong correlation in all four regions (see Figure~\ref{Fig_3}), with Pearson coefficients of ($r=0.88$), 0.92, 0.94, and 0.92 for the north, east, south, and west, respectively. In contrast, the radio-optical correlation varies substantially, from ($r=0.92$) in the north to ($r=0.36$) in the south. A similar trend is found for the X-ray–optical profiles, with the weakest correlation in the south ($r=0.31$). The peak radii also show significant azimuthal variations. In particular, the southern region is distinctive, with the radio and X-ray profiles peaking at approximately 31.6 and 46.9\,arcsec, respectively, while the optical emission peaks within the innermost radial region. These results indicate a strong radial correlation between the radio and X-ray emission across the remnant, whereas the optical emission exhibits a more pronounced directional dependence, particularly in the southern region.  

\begin{figure}
\begin{center}
\includegraphics[width=0.95\columnwidth,trim=0 0 0 0,clip]{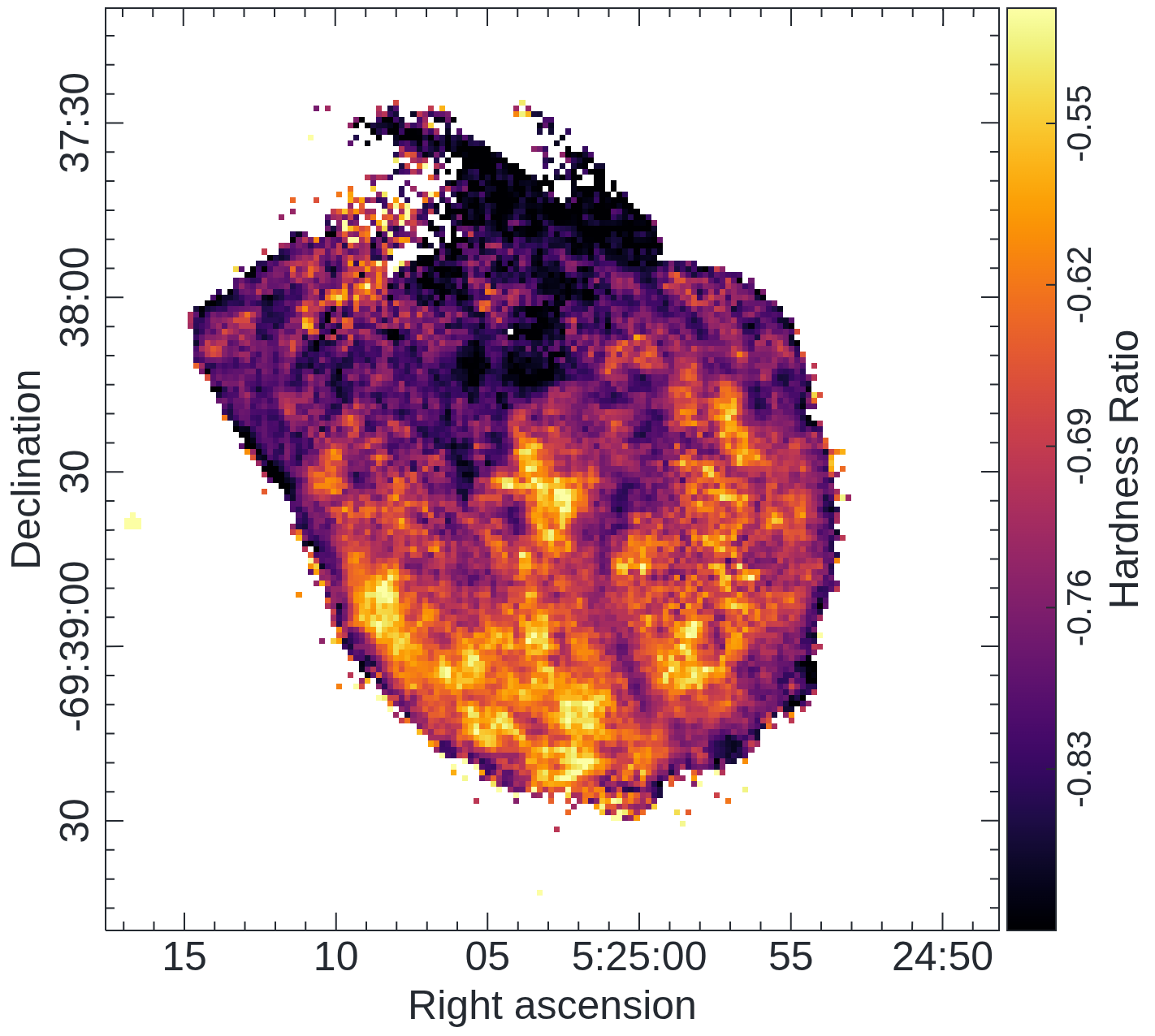}
\caption{Hardness ratio map between the 0.3--1.3 and 1.3--3.0\,keV bands obtained from \cxo\ data.}\label{Fig_2}
\end{center}
\end{figure}

The strong spatial correlation between the radio and X-ray emission may suggest a contribution from non-thermal X-ray emission. To investigate spatial variations in the X-ray spectral properties, we constructed soft- and hard-band images from the \cxo\ data in the 0.3--1.3 and 1.3--3.0\,keV energy bands, respectively. The two images were regridded to a common spatial grid, and the hardness ratio (HR) was calculated on a pixel-by-pixel basis as $\mathrm{HR}=(H-S)/(H+S)$, where $H$ and $S$ denote the hard- and soft-band count rates, respectively. The resulting hardness-ratio map shows that the X-ray emission is predominantly soft across the remnant (Figure~\ref{Fig_2}). The mean HR for the entire remnant is $\sim-0.75$, while the central and lower-left regions exhibit higher values of $\sim-0.65$. These higher HR values indicate a relatively enhanced contribution from the hard-band (1.3--3.0\,keV) emission in these regions compared with the remnant as a whole. Nevertheless, the negative HR values indicate that the X-ray emission remains dominated by the soft-band (0.3--1.3\,keV) component.

 Interestingly, the optical emissions lie outside both of radio and X-ray emissions (see Figure~\ref{Fig1}). This multi-frequency morphology is consistent with the expected shock structure for younger \acp{SNR}. The radio-continuum emission likely traces the synchrotron radiation from shock-accelerated electrons in the interior; the X-ray emission traces the hot, shocked plasma, both in the interior and at the outer shock front; and the \OIII\ optical emission extends further into the cooling zones beyond the shock. This outer diffuse optical emission, particularly obvious on the north-western rim and on the western side of the remnant, likely arises from ambient material which has been photoionised by the fast \ac{SNR} shock, as identified by \citet{1996AJ....112..509M}. Interestingly, the brightest \OIII\ arc is visible in the north-west, where the radio-continuum is faintest and most diffuse. This anti-correlation may suggest spatially varying shock structures, potentially due to the northern side expanding into a lower density region, where the ambient material is not dense enough for strong synchrotron emission but is still experiencing strong collisional ionisation or photoionisation, generating the \OIII\ emission. This scenario is consistent with the interpretation that \snr\ is interacting with a molecular cloud structure in the south~\citep{2020ApJ...902...53S}, thus creating an asymmetric density structure. 


\subsection{Polarisation, Rotation Measure, and Magnetic Field Strength } \label{pol}
The fractional polarisation for this \ac{SNR} was calculated using Equation~\ref{eq} where $P$ is the average fractional polarisation, and $S_{Q}$, $S_{U}$, and $S_{I}$ are integrated intensities for the $Q$, $U$, and $I$ Stokes parameters, respectively. 

\begin{equation}
\label{eq}
P=\frac{\sqrt{S^2_Q+S^2_U}}{S_I},
\end{equation}

\begin{figure*}
\centering





\includegraphics[width=0.95\linewidth]{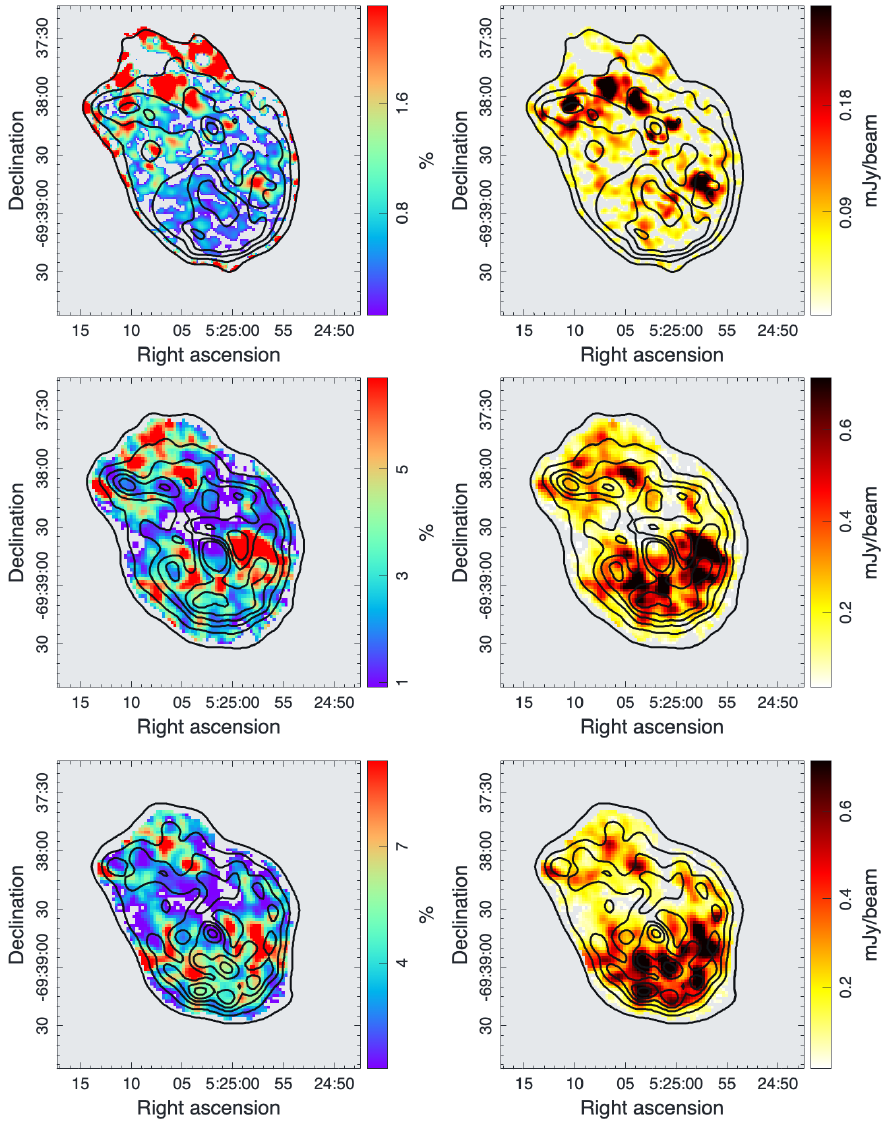}

\caption{\textbf{Left:} Fractional polarisation \textcolor{red} images of \snr\ 
at 2100\,MHz (top), 5500\,MHz (middle), and 9000\,MHz (bottom). 
The bar on the right side represents the fractional polarisation values in percentages. \textbf{Right:} Polarisation intensity maps of \snr\ at 2100\,MHz (top), 5500\,MHz (middle), and 9000\,MHz (bottom) with intensity image contour lines overlaid. The contour levels for both images are: 1, 5, 10, 15, and 20\,mJy\,beam$^{-1}$ for 2100\,MHz; 2.5, 5, 7.5, 10.0, 12.5, and 15.0\,mJy\,beam$^{-1}$ for 5500\,MHz; and 2, 4, 6, 8, 10, and 12\,mJy\,beam$^{-1}$ for 9000\,MHz. The colour bar represents gradients of polarisation intensity in mJy\,beam$^{-1}$.} 
\label{fig:pol.5500}
\end{figure*}

The fractional polarisation from \snr\ appears prominent at all frequencies as expected from a young \ac{SNR}. Fractional polarisation maps at 2100, 5500, and 9000\,MHz, as well as polarisation intensity maps at these frequencies, are presented in Figure~\ref{fig:pol.5500}. The average fractional polarisation values are $1.2\pm0.9$\,per\,cent, $3\pm1$\,per\,cent, and $4\pm1$\,per\,cent for 2100, 5500, and 9000\,MHz respectively. These values are slightly lower than the estimates reported by \cite{1995AJ....109..200D}, but are consistent with those found for young \acp{SNR} in the \ac{LMC} \citep[e.g.][]{2019Ap&SS.364..204A,2025PASA...42...69A}.

From the MeerKAT analysis, the Faraday Depth spectra across \snr\ reveal complex, overlapping structure with emission distributed primarily from approximately --180 to +200\,rad\,m$^{-2}$. A spectrum averaged over the entire \ac{SNR} shows a broad, double-peaked structure, with peaks near $\phi\approx-30$ and +70\,rad\,m$^{-2}$. Individual pixel spectra show systematic spatial variation in the relative contributions of these components, and this variation is the reason that the moment maps were used as opposed to the simpler peak-finding method. As the moment maps are integrated over the entire Faraday Depth, they are better able to characterise a broader spectral structure.

\begin{figure*}
\centering

\includegraphics[width=\textwidth]{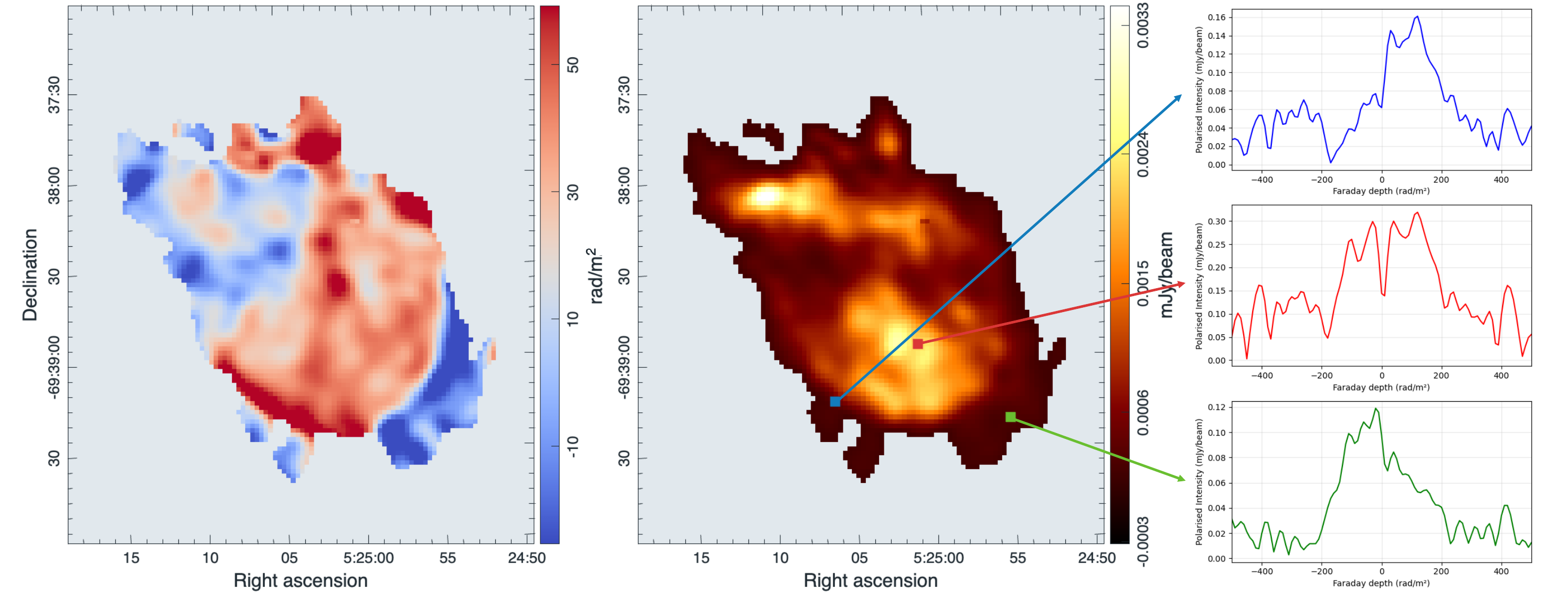}

\caption{Results of rotation measure synthesis performed on \snr\ using MeerKAT radio polarisation data over the Faraday Depth range -500 to +500 rad m$^{-2}$. The images are cut at a 10$\sigma$ local noise level. The first moment map (M$_1$, \textbf{left}) shows the intensity-weighted mean Faraday depth over the Faraday Depth range. The zeroth moment map (M$_0$, \textbf{middle}) shows the total polarised intensity integrated over the Faraday Depth range. \snr\ displays spatial variation in the Faraday Depth spectrum, with three distinct components represented by the coloured dots: south-west rim (blue dot), interior (red dot), and south-east outer region (green dot). Three example Faraday Depth spectra are shown as insets (\textbf{right}) showing the variation in these different regions, indicated by the colours and corresponding arrows.}
\label{fig:MeerKATpol}
\end{figure*}

The results are presented in Figure~\ref{fig:MeerKATpol} with both the M$_0$ and M$_1$ images cut at a 10$\sigma$ noise level. The M$_0$ map (middle panel) shows an asymmetric morphology with bright, well-defined emission in the south and a brighter arc of emission along the north side. The emission extends to the north-west, with an additional patch of emission visible in the south-west, outside of what appears to be the main \ac{SNR} structure. 

The M$_1$ map (left panel) shows clear spatial structure in the Faraday depth distribution. The south-western and north-eastern rims exhibit the most positive values (up to $\sim$80\,rad\,m$^{-2}$), while the interior shows intermediate positive values ($\sim$30\,rad\,m$^{-2}$). The north-western edge of the remnant shows a less regular structure, with patches of positive and negative M$_1$ values, ranging from $\sim-20$ to +20\,rad\,m$^{-2}$. The south-western patch of emission shows distinctly negative rotation measure (up to $\sim$--40\,rad\,m$^{-2}$). Although fainter than the main \ac{SNR} structure, this feature is detected at $>$10$\sigma$ significance and so is likely real emission.

Representative Faraday Depth spectra from three regions (Figure~\ref{fig:MeerKATpol}, right panel) demonstrate the spatial variation. The south-eastern rim (blue, top inset) shows a clear peak weighted toward positive values, consistent with the high positive M$_1$ values in this region. The interior spectrum (red, middle inset) shows a broad double-peaked structure, representative of what is seen in the whole \ac{SNR} averaged spectrum, with contributions from both the positive and negative Faraday Depth components. The south-western outer emission (green, bottom inset) shows structure weighted towards the negative Faraday Depths, as shown by the negative M$_1$ values in this region. This systematic variation indicates there are multiple magnetised regions along the line of sight with spatially varying contributions. 

This complex Faraday structure likely arises from different magnetised regions with partially varying magnetic field geometry. The positive M$_1$ values at portions of the rim (southeast and northwest) show emission from regions where the line-of-sight magnetic field component is directed towards us. These rim regions are viewed tangentially through the shell with minimal interior contribution. Therefore, these regions likely represent the outer shock front, and so could reflect compression and amplification of the ambient magnetic field, which was oriented toward us prior to the shock passage.

The negative component is most prominent in the south-western emission region, which is predominantly located just outside of the \ac{SNR} shell as seen in radio-continuum. The intermediate positive M$_1$ values in the interior are a result of a combination of both positive and negative components along the same line of sight, as evidenced by the double-peaked Faraday Depth spectra (Figure~\ref{fig:MeerKATpol}, middle panel and middle inset of right panel). 

The previous ALMA CO observations of \citet{2020ApJ...902...53S} showed a molecular cloud to the south of \snr\ which may be contributing to one of these Faraday Depth components. A possible scenario for the observed values is that the molecular cloud has a magnetic field oriented away from us, and so is the origin of the negative component. In this scenario, the intermediate values would then arise from sightlines passing through both the molecular cloud (for example, if it is also in front of \snr) and the \ac{SNR} shell itself. This would naturally result in an intermediate value as the weighted average of the positive and negative components.

Alternatively, if there is active interaction with the molecular cloud, then this may introduce magnetic turbulence into the \ac{SNR} interior. Shock-cloud interaction could create a complex interior magnetic field geometry with regions of opposing magnetic field orientation (e.g. compressed or twisted field lines), potentially producing the observed positive and negative components. Another possibility is that there may be an unrelated foreground rotation measure component along the sight (Galactic of \ac{LMC} based) which is responsible for one of the observed components, making it unrelated to \snr\ itself. 

\begin{figure}[t]
 \begin{center}
\includegraphics[width=\columnwidth,trim=25 5 10 20,clip]{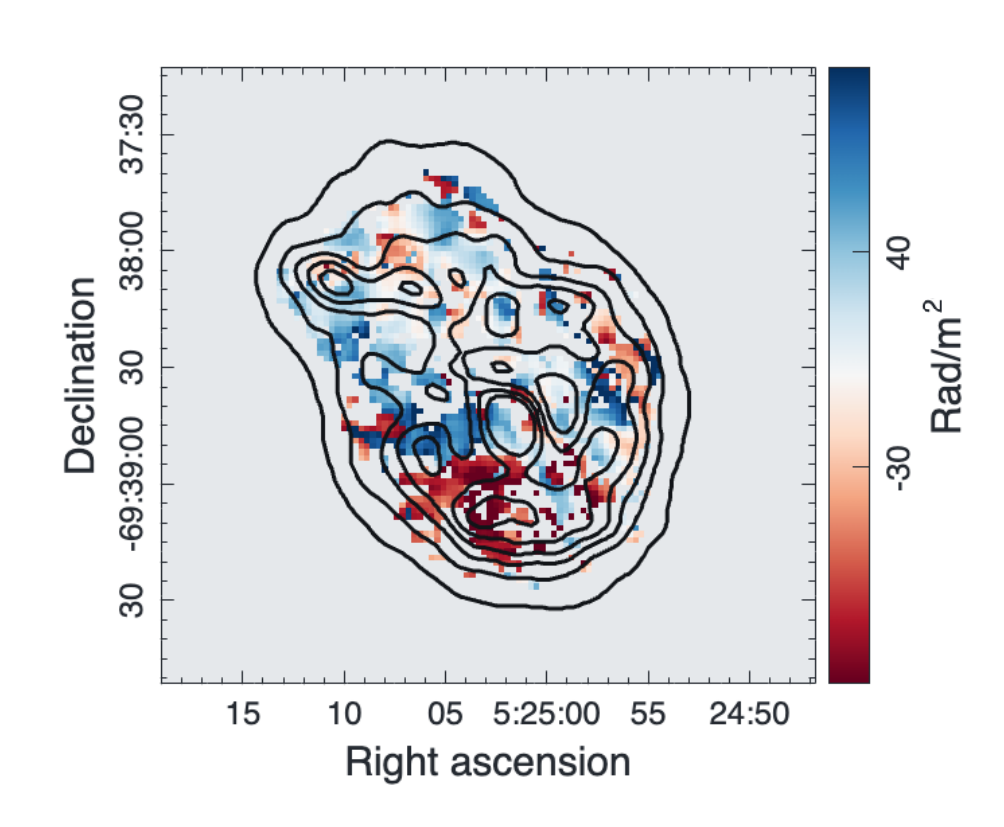}
 \caption{$RM$ map of \snr\ (\ac{ATCA}; 2100, 5500 and 9000\,MHz). The contour lines are from \ac{ATCA} 5500\,MHz image (2.5, 5, 7.5, 10.0, 12.5, and 15.0\,mJy\,beam$^{-1}$). The colour bar on the right side represents the gradient rotation measure in rad\,m$^{-2}$. }\label{RM}
\end{center}
 \end{figure}

\begin{figure}[htbp]
 \begin{center}
\includegraphics[width=\columnwidth,trim=25 5 10 20,clip]{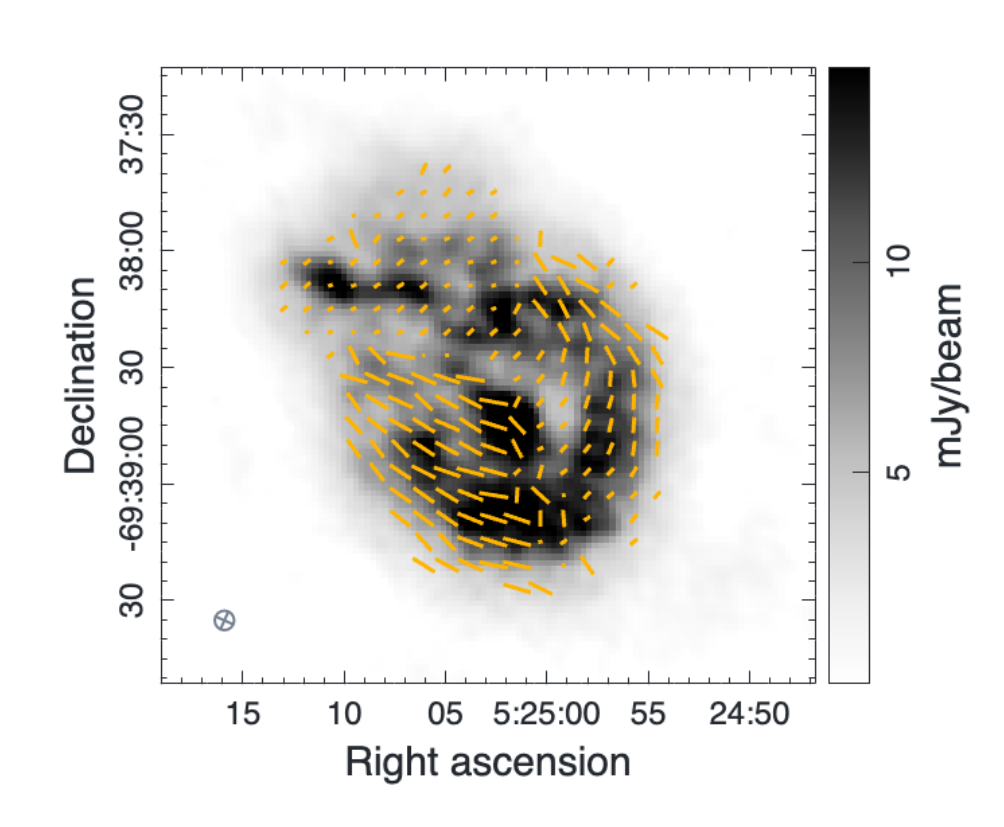}
 \caption{Magnetic field direction vectors of \snr\ overlaid on \ac{ATCA} intensity image at 5500\,MHz. The circle in the lower left corner represents a synthesised beam of $5\times5$\,\,arcsec$^{2}$ and the bar on the right side represents the gray scale gradients for \ac{ATCA} image in mJy\,beam$^{-1}$.}
\label{MDF}
\end{center}
 \end{figure}

The rotation measure ($RM$) of \snr\ was also calculated using position angle measurements associated with fractional polarisation maps at 2100, 5500, and 9000\,MHz, then applying \textsc{miriad} task \textsc{imrm}. $RM$ values with an error $\geq$~120\,rad\,m$^{-2}$ were flagged. The resulting $RM$ map is shown in Figure~\ref{RM} with an average value of $2\pm27$\,rad\,m$^{-2}$ and a maximum value of $338\pm70$\,rad\,m$^{-2}$ located on the western side of the remnant. Once the $RM$ had been determined, the vectors were rotated back to their intrinsic (zero wavelength) values. In Figure~\ref{MDF} an additional 90$^\circ$ was added to the electric vectors so that the vectors represent the direction of magnetic fields within the remnant. These vectors are oriented in a radial direction (see Figure~\ref{MDF}), as expected from a young \ac{SNR}.

The magnetic field strength can be estimated using the equation for Faraday depth\footnote{This assumes only one source along the line of sight with no internal Faraday rotation; the Faraday depth approximates the rotation measure of all wavelengths.}:

\begin{equation}
    \label{eq3}
    RM=811.9\int_{0}^{L}n_{e}B_{\parallel}dl ,
\end{equation}

\noindent where $RM$ is the rotation measure in rad\,m$^{-2}$, $n_e$ is electron density in cm$^{-3}$, $B_{\parallel}$ is line of sight magnetic field strength in $\mu$G, and $L$ is  path length through the Faraday rotating medium in kpc \citep{2004JKAS...37..337C}. 

Using our maximum value of $RM$ (338\,rad\,m$^{-2}$), an \ac{SNR} thickness ($L$) of the compressed shell of $\sim$12\,pc, and a $n_e$ value of 5\,cm$^{-3}$ \citep{1995AJ....109..200D}, we obtain a magnetic field strength of $\sim$7.9\,$\mu$G for the west side of the remnant. This result is higher than the value previously reported by \cite{1995AJ....109..200D}.

The equipartition model\footnote{Online calculator is found at \url{http://poincare.matf.bg.ac.rs/~arbo/eqp}} can also be used to estimate the magnetic field strength for \snr. This method uses modelling and simple parameters to estimate the intrinsic magnetic field strength and the energy contained in the magnetic field and \ac{CR} particles using radio synchrotron emission \citep{2012ApJ...746...79A,2013ApJ...777...31A,2018ApJ...855...59U}.
 
This approach is purely analytical, described as a rough, order-of-magnitude estimate because of assumptions used in analytical derivations and errors in the determination of distance, angular diameter, spectral index, a filling factor, and flux density, tailored especially for the magnetic field strength in \acp{SNR}. \citet{2012ApJ...746...79A,2013ApJ...777...31A} and \citet{2018ApJ...855...59U} present two models; the difference is in the assumption of whether there is equipartition, precisely constant partition with \acp{CR} or only \ac{CR} electrons. \citet{2018ApJ...855...59U} showed the latter type of equipartition is a better assumption than the former. 

Using the \citet{2018ApJ...855...59U} model\footnote{We use: $\alpha=0.74$, $\theta$~=~2\,arcmin, $\kappa=0$, S$_{\rm 2.1\,GHz}$~=~3.05\,Jy, and f~=~0.25.}, the mean constant partition field over the entire remnant is $\sim$90\,$\mu$G with an estimated minimum energy of E$_{\rm min}$~=~3.3$\times10^{50}$\,erg. However, using the \citet{2012ApJ...746...79A} earlier model, which assumes \acp{CR} are composed of electrons, protons, and ions, the estimate is $\sim$125\,$\mu$G, with a minimum explosion energy of E$_{\rm min}$~=~1.1$\times10^{51}$\,erg.

Finally, if the magnetic field estimate is $\sim$90\,$\mu$G, and given that \snr\ is a young \ac{SNR}, this indicates nonlinear effects in particle acceleration, which produce magnetic field amplification, not only compression on the shock wave. Compression on the shock wave would make the field $\sim$20\,$\mu$G, compared to our estimate of $\sim$90\,$\mu$G, which is made by magnetic field amplification produced by the non-linear particle acceleration effects \citep{2014Ap&SS.354..541U,2020NatAs...4..910U,2022PASP..134f1001U}.

\subsection{Spectral Index}

\begin{figure}
 \begin{center}
\includegraphics[width=\columnwidth,trim=25 5 10 20,clip]{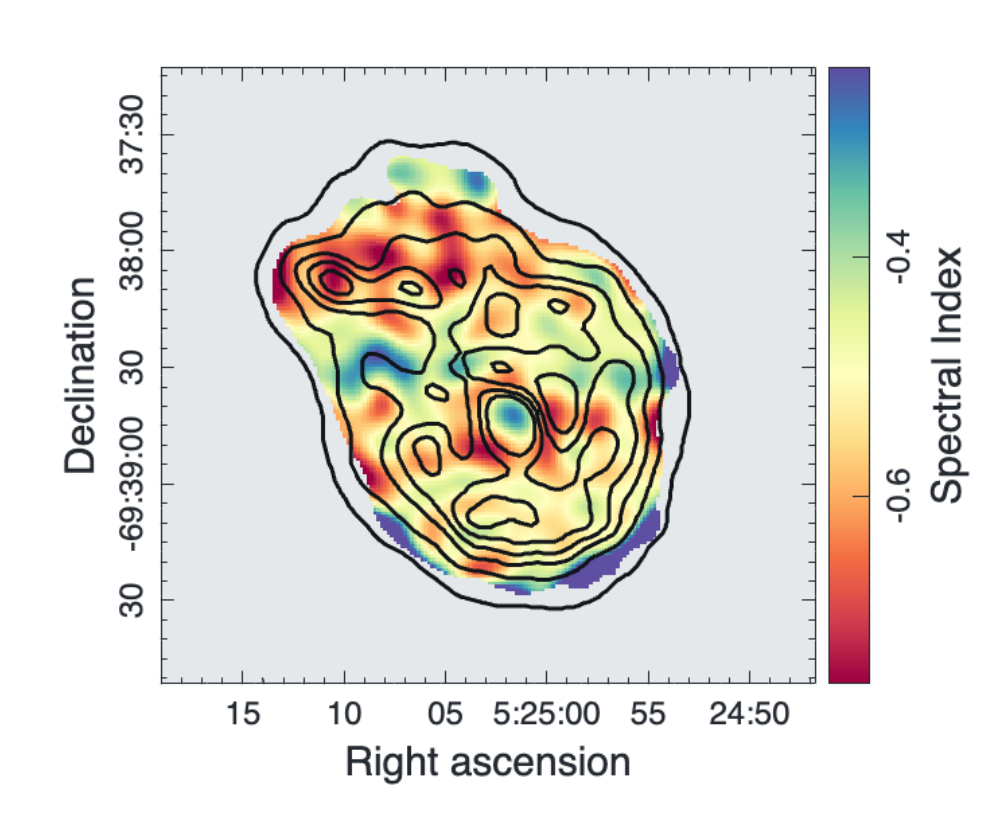}
 \caption{Spectral index map of \ac{SNR} \snr\ between MeerKAT (1295\,MHz) and \ac{ATCA} (2100, 5500, and 9000\,MHz) images. The contour lines are from \ac{ATCA} 5500\,MHz image (2.5, 5, 7.5, 10.0, 12.5, and 15.0\,mJy\,beam$^{-1}$). The colour bar on the right side represents the gradients of the spectral index. }\label{Fig3}
\end{center}
 \end{figure}

The radio spectra of this \ac{SNR} can often be described as a 'pure power-law of frequency'. A spectral index map for \snr\ was produced using the MeerKAT (1295\,MHz) and \ac{ATCA} (2100, 5500, and 9000\,MHz) images. All images were first re-gridded to the finest pixel size ($0.6\times0.6$\,arcsec$^{2}$) using the \textsc{miriad} task \textsc{regrid}. The images were then smoothed to the lowest resolution ($8\times8$\,\,arcsec$^{2}$) using the \textsc{miriad} task \textsc{convol}. Finally, the \textsc{miriad} task \textsc{maths} was applied to create the spectral index map as shown in Figure~\ref{Fig3}. The average spectral index across the remnant is $\alpha \sim-0.5$.

\begin{figure}[t!]
 \begin{center}
\includegraphics[width=\columnwidth,clip]{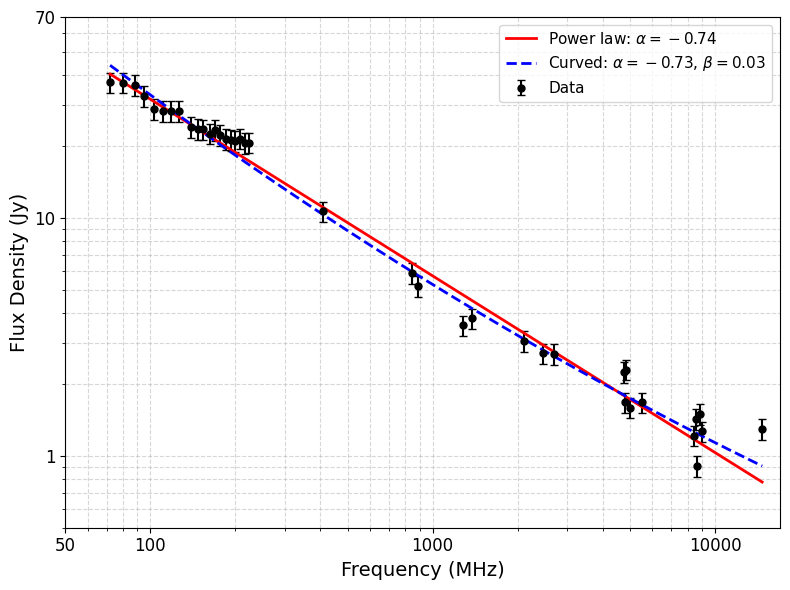}\\
 \caption{Radio continuum spectrum of \snr. The flux density uncertainties are too small to be visible.}\label{Fig4}
\end{center}
 \end{figure}

In Figure~\ref{Fig4}, we plot integrated flux densities at a wide range of frequencies to estimate the \snr\ overall spectral index. We used the \textsc{miriad} \citep{1995ASPC...77..433S} task \textsc{imfit} to extract a total integrated flux density from all available radio-continuum observations of \snr\ listed in Table~\ref{tab2}. The observations are combined with flux density measurements from the Murchison Widefield Array~\citep[MWA;][]{2018MNRAS.480.2743F}\acused{MWA}, Australian Square Kilometre Array Pathfinder~\citep[ASKAP;][]{2021MNRAS.506.3540P}\acused{ASKAP}, Molonglo Observatory Synthesis Telescope~\citep[MOST;][]{2021MNRAS.507.2885F}\acused{MOST}, pre \ac{CABB} \ac{ATCA} \citep{1995AJ....109..200D}, Parkes Telescope \citep{2017ApJS..230....2B}, MeerKAT, and \ac{CABB} \ac{ATCA} (Table~\ref{tab2}). We measured the \ac{MWA} flux density for each sub-band (72--223\,MHz). The red line represents the best power-law weighted least-squares fit\footnote{No intrinsic scatter is considered in estimating spectral index and thus, the uncertainty is very likely underestimated.} with the spatially integrated spectral index \mbox{$\langle\alpha\rangle= -0.74 \pm 0.01$}, consistent with the previous estimate by \cite{1995AJ....109..200D}. 


\begin{table}[th!]
\caption{Flux density measurements of \snr.}
\centering
\begin{tabular}{@{}cclcc@{}}
\hline\hline
Frequency &    Fluxes &     Telescope & Reference\\
(MHz)     &     (Jy)  &            \\
 \hline%
72        &     37.19         &\ac{MWA}        &  This work\\
80        &     37.12         &\ac{MWA}        & This work\\
88        &    36.28          &\ac{MWA}        &  This work\\
95        &    32.65          &\ac{MWA}        & This work\\
103       &    28.83          &\ac{MWA}        & This work\\
111       &    28.22          &\ac{MWA}        &  This work\\
118       &    28.15          &\ac{MWA}        &  This work\\
126       &    28.18          &\ac{MWA}        &  This work\\
139       &    24.17          &\ac{MWA}        &  This work\\
147       &    23.72          &\ac{MWA}        &  This work\\
154       &    23.57          &\ac{MWA}        &  This work\\
162       &    22.66          &\ac{MWA}        &  This work\\
170       &    23.53          &\ac{MWA}        & This work\\
177       &    22.28          &\ac{MWA}        &  This work\\
185       &    21.54          &\ac{MWA}        &  This work\\
193       &    21.27          &\ac{MWA}        &  This work\\
200       &    21.06          &\ac{MWA}        &  This work\\
208       &    21.60          &\ac{MWA}        &  This work\\
216       &    20.75          &\ac{MWA}        &  This work\\
223       &    20.78          &\ac{MWA}        & This work\\
408       &    10.67          &\ac{MOST}       &  \cite{2021MNRAS.507.2885F}\\
843       &    5.89           &\ac{MOST}       &  \cite{2021MNRAS.507.2885F}\\
888       &    5.18           &\ac{ASKAP}      &  \cite{2021MNRAS.506.3540P}\\
1295      &    3.54           &  MeerKAT       &  This work          \\
1377      &    3.79           & Parkes         &  \cite{2017ApJS..230....2B}  \\
2100      &    3.05           &\ac{ATCA}       &  This work\\
2450      &    2.70           & Parkes         & \cite{2017ApJS..230....2B}   \\
2700      &    2.68           &  Parkes        &   \cite{2017ApJS..230....2B}  \\
4750      &    2.25           &   Parkes       &  \cite{2017ApJS..230....2B}   \\
4800      &    1.68           & \acs{ATCA}     &  \cite{1995AJ....109..200D}   \\
4850      &    2.31           &  Parkes        &  \cite{2017ApJS..230....2B}    \\
5000      &    1.60           &  Parkes        &    \cite{2017ApJS..230....2B}   \\
5500      &    1.68           &\ac{ATCA}       &  This work\\
8400      &    1.22           &    Parkes      &    \cite{2017ApJS..230....2B}         \\
8550      &    1.43           &   Parkes       &    \cite{2017ApJS..230....2B}         \\
8640      &    0.91           &  \acs{ATCA}    &   \cite{1995AJ....109..200D}           \\
8873      &    1.50           &    Parkes      &    \cite{2017ApJS..230....2B}          \\
9000      &    1.27           &\ac{ATCA}       &  This work\\
14700     &    1.30           &    Parkes      &    \cite{2017ApJS..230....2B}       \\
\hline

\hline\hline
\end{tabular}
\label{tab2}
\end{table}

For cross-checking and consistency, we used \textsc{aegean} \citep{hancock2018} and found no significant difference in integrated flux density estimates. 
Namely, we measured \snr\ local background noise (1$\sigma$) and carefully selected the exact area of the \ac{SNR} which also excludes all obvious unrelated point sources. We then estimated the sum of all brightnesses above 5$\sigma$ of each individual pixel within that area and converted it to \ac{SNR} integrated flux density following \citet[][Equation~24]{1966ARA&A...4...77F}. 
We also estimate that the corresponding radio flux density errors are below 10\,per\,cent as examined in our previous work \citep{2022MNRAS.512..265F,2023MNRAS.518.2574B,2025PASA...42...69A}.

Taking all these factors into account, the spectrum in \snr\ appears to exhibit a concave-up (Figure~\ref{Fig4}, blue dashed curve), which is characteristic of nonlinear effects in young \acp{SNR} where particle acceleration is highly efficient and produces a significant population of \acp{CR}. The accelerated \acp{CR} can modify the shock structure, resulting in a concave-up radio spectrum \citep{2014Ap&SS.354..541U}.

\section{Conclusion} \label{con}

In this work, we have presented a detailed radio continuum study of the \ac{LMC} \ac{SNR} \snr. Our results reveal a strong spatial correlation between the radio and X-ray emission across the central and left regions of the remnant, as well as along the right section of the shell, highlighting the close link between thermal and non-thermal processes. The remnant shows significant polarisation, with average fractional polarisation of $1.2\pm0.9$\,per\,cent, $3\pm1$\,per\,cent, and $4\pm1$\,per\,cent at 2100, 5500, and 9000\,MHz, respectively, indicating an ordered magnetic field structure in parts of the shell.

The Faraday depth structure, derived from MeerKAT data, is complex and suggests the presence of multiple magnetised components along the line of sight. In contrast, the $RM$, obtained from \ac{ATCA} observations, has an average value of $2\pm27$\,rad\,m$^{-2}$ and reaches a maximum of $338\pm70$\,rad\,m$^{-2}$, further supporting a structured and turbulent magneto-ionic environment. From a constant partition model, we estimate a magnetic field strength of $\sim$90\,$\mu$G and a minimum energy of E$_{\rm min}$~=~3.3$\times10^{50}$\,erg, confirming that \snr\ is a highly energetic remnant. Finally, the steep radio spectral index of $-0.74\pm0.01$ is consistent with synchrotron emission from shock-accelerated electrons in young \acp{SNR}.

Together, these results provide a comprehensive picture of the magnetic field properties, energetics, and particle acceleration processes operating in \snr.

\section*{Acknowledgments}

The Australian Compact Array is part of the Australian Telescope which is funded by the Commonwealth of Australia for operation as National Facility managed by \ac{CSIRO}. This paper includes archived data obtained through the \ac{ATOA} (\url{http://atoa.atnf.csiro.au}). We used the \textsc{karma} and \textsc{miriad} software packages developed by the \ac{ATNF}. We thank the referee for the valuable comments and constructive suggestions, which significantly improved the manuscript. 

\bibliographystyle{apsrev4-1}

\bibliography{oja_template}




\end{document}